\documentclass[12pt]{article}

\makeatletter
\renewcommand\@biblabel[1]{#1.}
\makeatother

\usepackage{pstricks,pst-node,graphicx}
\usepackage{amssymb,amsmath,mathrsfs,setspace}
\usepackage{caption}

\newtheorem{remark}{Remark}[section]
\newtheorem{lemma}{Lemma}[section]
\newtheorem{proposition}{Proposition}[section]
\newtheorem{theorem}{Theorem}[section]

\newcommand{\be}{\begin{equation}}
\newcommand{\ee}{\end{equation}}
\newcommand{\eea}{\end{eqnarray}}
\newcommand{\bea}{\begin{eqnarray}}

\newcounter{secnum}[section]
\begin{document}

\title{{\bf Next steps in understanding the asymptotics of $3d$ quantum gravity}
\footnotetext{\newline{\bf Key words and phrases}: quantum gravity, coloured hard-dimers, generating function, central limit theorem, special functions.
\newline
{\it Mathematics Subject Classification (2010)}: 60B20, 60F05, 05A15, 60C05, 83C27}}

\author{{\bf Maria Simonetta Bernabei$\hbox{}^{1}$} \\ and \\
{\bf Horst Thaler$\hbox{}^{2}$} \\[1ex]
$^1$ School of Science and Technology, Department of Mathematics,   \\
University of Camerino,
Via Madonna delle Carceri 9,\\
I--62032, Camerino (MC), Italy\\
$^2$ School of Architecture and Design, University of Camerino, \\
Viale della Rimembranza, I--63100, Ascoli Piceno (AP), Italy \\
{\small simona.bernabei@unicam.it, horst.thaler@unicam.it}}

\date{}

\maketitle

\abstract{{Based on a combinatorial approach and random matrix theory, we show a central limit theorem that gives important insight into causally triangulated $3d$ quantum gravity. 
}}

\section{Introduction}
 \setcounter{secnum}{\value{section}
 \setcounter{equation}{0}
 \renewcommand{\theequation}{\mbox{\arabic{secnum}.\arabic{equation}}}}

We examine here a model of quantum gravity that was invented by Benedetti, Loll and Zamponi (BeLoZa) in the framework of causally triangulated $(2+1)$-dimensional quantum gravity, see \cite{BeLoZa07}. These authors have initiated an important discussion about the characteristics of this model. We analysed the model further and have discovered that the continuum limit can be grasped by essentially two parameters that stem from a central limit theorem involved in it. 

The particular feature of the model used in \cite{BeLoZa07} is the usage of special triangulations of space-time. The configurations of each time slice are made of triangulations of prisms whose basis itself forms a two-dimensional triangulation.
This makes the generating function of the one-step propagator expressible in terms of coloured hard-dimers.

\begin{figure}[h]
\captionsetup{width=0.8\textwidth}
\begin{center}
\psset{xunit=1.3cm,yunit=0.5cm}
\begin{pspicture}(0,0)(7,1.2)
\psdots[dotstyle=*,linecolor=red,dotscale=1.2](0.5,0.5)(2.0,0.5)(3,0.5)(3.5,0.5)(4.5,0.5)(6.5,0.5)
\psdots[dotstyle=*,linecolor=blue,dotscale=1.2](1,0.5)(1.5,0.5)(2.5,0.5)(4,0.5)(5,0.5)(5.5,0.5)(6,0.5)
\psline[linecolor=red,linewidth=1pt](0.5,0.5)(2,0.5)
\psline[linecolor=red,linewidth=1pt](3,0.5)(3.5,0.5)
\psline[linecolor=blue,linewidth=1pt](4,0.5)(5,0.5)
\end{pspicture}
\caption{\label{hdimer}{\small A hdc $(N=13)$ with two red dimers, one blue dimer and four single points.}}
\label{fig1}
\end{center}
\end{figure}
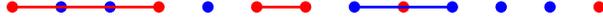

What are coloured hard-dimers: Given a sequence $\xi_N$ of length $N$ of blue and red sites on the one-dimensional lattice $\mathbb{Z}$, one defines a dimer to be an edge connecting two nearest sites of the same colour, which characterizes the dimer colour. A sequence of coloured and non-overlapping dimers (``hardness condition") in turn yields a ``coloured hard-dimer" (CHD), in the following also termed hard-dimer configuration (hdc). In figure \ref{fig1} an example of a coloured hard-dimer is given.
\\[-2ex]
Our goal is to study the asymptotics of the one-step propagator in the $(2+1)$-dimensional case, more precisely, of its generating function $Z(x,y,\Delta t=1)$.
By an inversion formula (see \cite[eq.(19)]{BeLoZa07}), $Z$ can be expressed in terms of hard-dimer configurations (hdc's). This inversion formula translates the generating function $Z_{\xi_N}$, which is related to tower geometries for fixed $\xi_N$, into one-dimensional objects $Z_{\textnormal{hdc's on}\: \xi_N}$, giving 
\bea
\label{nr21}
Z(x,y,\Delta t=1) & =& \sum\limits_{N} e^{-\gamma N} \sum\limits_{\xi_N} Z_{\xi_N}(u,v,w)  \nonumber \\
    & =& \sum\limits_{N} e^{-\gamma N} \sum\limits_{\xi_N} \displaystyle{\frac{1}{Z_{\text{hdc's on}\: \xi_N}(-u,-v,w)}}\nonumber \\
    &=& \sum\limits_{N} e^{(\ln 2-\gamma) N} \left< \displaystyle{\frac{1}{Z_{\text{hdc's on}\: \xi_N}(-u,-v,w)}} \right>,
\eea
where $Z_{\textnormal{hdc's on}\: \xi_N}$ is the combinatorial generating function associated with the configuration $\xi_N$:
 \be
 \label{nr22}
 Z_{\textnormal{hdc's on}\: \xi_N}(u,v,w) = \sum_{s_b,s_r,m} N_{\xi_N}(s_b,s_r,m) u^{s_b} v^{s_r} w^{m}
 \ee
and $N_{\xi_N}(s_b,s_r,m)$ is the number of hdc's on $\xi_N$ containing $s_b$ blue dimers, $s_r$ red dimers and $m$ mixed points (number of sites within each dimer).
The configuration space carries a uniform probability measure where each configuration $\xi_N$ has probability $1/2^N$ to be chosen. 
The expectation $\langle\cdot \rangle$ refers to this probability space. 

The central limit theorem entails an asymptotic behviour of the form
\be
\label{nr23}
\left<\frac{1}{Z_{\textnormal{hdc's on}\: \xi_N}(-u,-v,w)}\right>\asymp e^{N \beta^2}\frac{1}{\langle Z_{\textnormal{hdc's on}\: \xi_N}(-u,-v,w)\rangle},
\ee
as $N \to \infty$, where $\beta^2$ denotes an asymptotic variance to be defined below. Relation (\ref{nr23}) supplies one basic ingredient to investigate the properties of (\ref{nr21}).

Introducing the variables $s=s_b+s_r$, i.e. the total number of dimers  and $k=N-s-m$,
which corresponds the total number of dimers and single points, we find 
\be
\label{nr05} 
\langle Z_{\textnormal{hdc's on}\: \xi_N}(-u,-v,w)\rangle = 1+
\sum\limits_{k=1}^{{N-1}} \; \sum\limits_{s=1}^{k\wedge(N-k)} \binom{k}{s} \binom{N-k-1}{s-1} \left (-\frac{u+v}{2w} \right )^s \left (\frac{w}{2}\right )^{N-k}.
\ee
Equality (\ref{nr05}) can be read off directly from formula (2.6) in \cite{BeTh10}, since 
$$
\langle Z_{\text{hdc's on}\: \xi_N}(-u,-v,w)\rangle = \frac{1}{2^N} \sum_{\xi_N} Z_{\text{hdc's on}\: \xi_N}(-u,-v,w)=\frac{1}{2^N}\tilde{F}_N(-u,-v,w).
$$
The combinatorial approach followed here will turn out to be crucial when proving, as we shall do in section 4, the following exact formula
\be\label{nr06}
\langle Z_{\textnormal{hdc's on}\: \xi_N}(-u,-v,w)\rangle =\frac{\tilde{C}}{S} \left (\frac{1+S+\frac{w}{2}}{2} \right )^{N-1},
\ee
where the constants $\tilde{C}\equiv \tilde{C}(u,v,w)$ and $S\equiv S(u,v,w)$ will be defined later.

The paper is organised as follows. In section 2 it is shown how the combinatorial generating functions can be written in terms of random matrices and the spectral properties of the latter are studied. In section 3 the central limit theorem is tackled, in particular the sufficient conditions for its validity are discussed in detail. In section 4 we give an explicit formula for the mean of the hard-dimer generating function (\ref{nr22}).
In section 5 we look at the implications our findings have for $3d$ quantum gravity and give further evidence that $(u_c,v_c,w_c)=(2/9,2/9,2/3)$ is the critical point where to perform the continuum limit.
\section{Representing $Z_{\textnormal{hdc's on}\: \xi_N}(-u,-v,w)$ in terms of random matrices}
 \setcounter{secnum}{\value{section}
 \setcounter{equation}{0}
 \renewcommand{\theequation}{\mbox{\arabic{secnum}.\arabic{equation}}}}

Using the results of a previous work, see \cite[Theorem 2.1, Remark 2.2, Proposition 3.1]{BeTh11},  one finds that $Z_{\text{hdc's on}\;\xi_N}(-u,-v,w)$ is nothing but the coefficient $c_{\mathbf{x}}$ of the recognizable series
$$
T=\sum_{\mathbf{x}}c_\mathbf{x}\mathbf{x},
$$
where
$$
c_\mathbf{x}=\sum_{i,j,k \in \mathbb{N}_0} m_{i,j,k}(\mathbf{x)} (-u)^i (-v)^j w^k,
$$
once the following identifications have been made: $\xi_N \equiv \mathbf{x}$, $(s_b,s_r,m)=(i,j,k)$  and $N_{\xi_N}(s_b,s_r,m)\equiv m_{i,j,k}(\mathbf{x})$.
Recognizable means that there is a representation of the monoid of words made of letters $B,R$ in terms of finite-dimensional matrices and two specific vectors $\lambda,\gamma$ so that
$$
c_{\mathbf{x}}= \langle\lambda, \text{(product of $B$'s and $R$'s corresponding to $\xi_N$)} \gamma\rangle_E\,.
$$
{\it Notation.} The symbol $\langle \cdot,\cdot \rangle_E$ stands for the standard inner product on $\mathbb{R}^d$ and should not be confused with the expectation $\langle \cdot,\cdot \rangle$ over some probability space, which is written without the subscript $E$.\\[1ex]
In our case the representation matrices are given as
\be
\label{nr25}
B=\left(
    \begin{array}{ccc}
      1 & 1 & 0 \\
      -u & 0 & 0 \\
      0 & 0 & w \\
    \end{array}\right ),
    \qquad
R=\left(
    \begin{array}{ccc}
      1 & 0 & 1 \\
      0 & w & 0 \\
      -v & 0 & 0 \\
    \end{array}
  \right),
  \ee
and the vectors are $\lambda=\gamma=(1,0,0)$.
Therefore we obtain
\be\label{randomproduct}
Z_{\text{hdc's on}\;\xi_N}(-u,-v,w)=\langle\lambda, \text{(product of $B$'s and $R$'s corresponding to $\xi_N$)} \gamma\rangle_E,
\ee
which in turn implies that 
\be\label{MeanZhdc}
\langle Z_{\text{hdc's on}\;\xi_N}(-u,-v,w)\rangle = \langle (\mathcal{S}_N)_{11} \rangle,
\ee
with $\mathcal{S}_N$ being a matrix-valued random variable. To be precise,
\be\label{SN}
\mathcal{S}_N=M_1 M_2\cdots M_N,
\ee
where each $M_i$ is a random variable assuming values $B$ or $R$ with probability $\frac{1}{2}$.
In this manner we have rewritten $Z_{\text{hdc's on}\;\xi_N}(-u,-v,w)$ as a matrix product and this fact will allow us to employ the central limit theorem from random matrix theory when studying the limit $N \rightarrow \infty$.

First we shall analyse the spectrum of the matrices $B, R, BR$ and $RB$.
Let $\mathrm{Gl}(3,\mathbb{R})$ be the set of invertible matrices with real entries and $M=\{R, B\}$.  
Note that $B,R \in \mathrm{Gl}(3,\mathbb{R})$. In fact, $\text{det}(B)=uw\neq0$, and $\text{det}(R)=vw\neq0$.
Let $\eta$ be the uniform distribution on $M$, so that $\eta(B)=\eta(R)=\frac{1}{2}$.
\begin{proposition}
The sum of the geometric multiplicities of each of the matrices $B, R, BR$ and $RB$ is at least $2$ and all the real eigenvalues are positive. Moreover, the sums of geometric multiplicities of each of the power matrices
$B^n, R^n, (BR)^n$ and $(RB)^n$, $n \geq 1$, are the same as those of $B, R, BR$ and $RB$, respectively.
\end{proposition}
\medskip
\noindent
{\bf Proof:} Let us consider the matrix $B$. The eigenvalues are
\be
\label{nr26}
\lambda_{1,2}=\frac{1 \pm \sqrt{1-4u}}{2},
\qquad
\lambda_{3}=w,
\ee
and the corresponding eigenvectors are
\be
\label{nr262}
\underline{e}_{1,2}=(-\frac{1 \pm \sqrt{1-4u}}{2u},1,0),
\qquad
\underline{e}_{3}=(0,0,1).
\ee

\noindent
For $u \neq \frac{1}{4}$, the set $\{ \underline{e}_{1}, \underline{e}_{2}, \underline{e}_{3}\}$ forms a basis of $\mathbb{R}^3$, so that it is also a basis for $B^n$, $n\geq1$.
Hence the assertion is valid in this case.

In the case where $u = \frac{1}{4}$, the sum of geometric multiplicities is $2$. By the Jordan normal form of the matrix $B$, the same is true for $B^n$. In fact, the Jordan normal form of $B$, for $u = \frac{1}{4}$ and $w \neq \frac{1}{2}$, is the following one
$$
J_B=\left(
\begin{array}{ccc}
\frac{1}{2} & 1 & 0 \\
0 & \frac{1}{2} & 0 \\
0 & 0 & w
\end{array}
\right).
$$
From the identity $B=PJ_BP^{-1}$, with matrix $P$ given by
$$P=\left(
\begin{array}{ccc}
-2 & -4  & 0 \\
1 & 0  & 0 \\
0 & 0 & 1 \\
\end{array}
\right),
$$
we deduce that $B^n=PJ^n_B P^{-1}$
$$
J^n_B=\left(
\begin{array}{ccc}
\frac{1}{2^n} & \frac{n}{2^{n-1}}  & 0 \\
0 & \frac{1}{2^n}  & 0 \\
0 & 0 & w^n \\
\end{array}
\right).
$$
Since the matrices $B^n$ and $J^n_B$ have the same number of linearly independent eigenvectors and the geometric multiplicity corresponding to eigenvalue $\frac{1}{2^n}$ is $1$, we find that the proposition holds true in this case.
In the last case when the matrix $B$ has a unique eigenvalue $\lambda =\frac{1}{2}$, i.e. for $u=\frac{1}{4}$ and $w=\frac{1}{2}$, the geometric multiplicity of the Jordan normal form is $2$, because
$$
J_B=\left(
\begin{array}{ccc}
\frac{1}{2} & 0 & 0 \\
0 & \frac{1}{2} & 1 \\
0 & 0 & \frac{1}{2} \\
\end{array}
\right),
$$
and hence the proposition also holds in this case.

For the matrix $R$ the proof of the statement is analogous. The eigenvalues of $R$ are
\be
\label{nr263}
\lambda_{1,2}=\frac{1 \pm \sqrt{1-4v}}{2},
\qquad
\lambda_{3}=w,
\ee
and the corresponding eigenvectors are
\be
\label{nr264}
\underline{e}_{1,2}=(-\frac{1 \pm \sqrt{1-4v}}{2v},0,1),
\qquad
\underline{e}_{3}=(0,1,0).
\ee
Note that for $u=v$, the matrices $B$ and $R$ have the same eigenvalues, but the eigenvectors are different.

Now consider the matrix
$$
BR=\left(
\begin{array}{ccc}
1 & w  & 1 \\
-u & 0  & -u \\
-vw & 0 & 0 \\
\end{array}
\right).
$$
The corresponding characteristic polynomial is
$\lambda^3-\lambda^2+w(u+v)\lambda -uvw^2=0$. From a qualitative study of the latter and taking into account that the physical parameters $u,v,w$ are real, we deduce that the eigenvalues of the matrix $BR$ can be real and positive, or complex. When the number of distinct eigenvalues is $3$,
we may argue as above to conclude that the sum of geometric multiplicities related to $(BR)^n$ is $3$.
Let us turn to the more interesting degenerate cases. First, consider the case of two distinct eigenvalues. This happens precisely when $\Delta=0$, where
$\Delta = \frac{q^2}{4}+\frac{p^3}{27}$, with
$p=w(u+v)-\frac{1}{3}\neq 0$ and $q=-uvw^2+\frac{w(u+v)}{3}-\frac{2}{27}\neq 0$.
It is easy to see that $\Delta =0$ if and only if
$$27 u^2 v^2 w^2 -(u+v)^2-18 uvw(u+v)+4uv+4w(u+v)^3=0$$ holds.
Note that the conjectured critical point $(u_c,v_c,w_c)=(\frac{2}{9}, \frac{2}{9}, \frac{2}{3})$ belongs to this degenerate case. The eigenspace corresponding to the eigenvalue $\lambda$ with algebraic multiplicity $2$ has dimension $1$. In fact, the corresponding eigenspace is given by $c\,(-\frac{\lambda}{vw}, -u(\frac{1}{\lambda}-\frac{1}{vw}),1)$, with $c\in\mathbb{R}$.
Arguing as above, by the Jordan normal form the sum of geometric multiplicities of $(BR)^n$ is $2$. The case where one has a unique eigenvalue with algebraic multiplicity $3$ is impossible, for this would imply $q=0 \Rightarrow p=0$. But there does not exist any $(u,v,w)$ such that $p=q=0$. The theorem is so proved for the matrix $BR$. The proof of the statement for $RB$ is analogous. Note that the matrix
$$
RB=\left(
                                                                                                                       \begin{array}{ccc}
                                                                                                                            1 & 1 & w \\
                                                                                                                            -uw & 0 & 0 \\
                                                                                                                            -v & -v & 0 \\
                                                                                                                          \end{array}
                                                                                                                        \right)
                                                                                                                        $$
                                                                                                                    has the same characteristic polynomial as $BR$.
                                                                                                                        \hfill{$\Box$}
\begin{remark}
From Proposition 2.1 it follows that
\be
\widehat{M}^n \underline{v}=\lambda \underline{v}\;\text{and $\lambda$ real}\;\text{implies}\; \lambda >0 \; \text{and} \;\;
\widehat{M} \underline{v} = \sqrt[n]{\lambda} \underline{v}
\ee
for any $\widehat{M}\in \{B, R, RB, BR\}$.
\end{remark}

\section{The central limit theorem}
 \setcounter{secnum}{\value{section}
 \setcounter{equation}{0}
 \renewcommand{\theequation}{\mbox{\arabic{secnum}.\arabic{equation}}}}

In light of equality (\ref{randomproduct}) it is advantageous to look at the asymptotic behaviour of  $Z_{\text{hdc's on}\;\xi_N}(-u,-v,w)$, as $N\rightarrow\infty$, from the perspective of random matrices.
Given a sequence $X_1, X_2,\ldots, X_n, \ldots$ of i.i.d. random matrices, distributed according to a distribution $\mu$ on $\mathrm{Gl}(d, \mathbb{R})$, $d\in \mathbb{N}$, and let $S_n=X_n X_{n-1} \cdots X_1$ be their product. Under suitable hypotheses a law of large numbers and a central limit theorem holds for the elements $\ln |\langle S_n x,y \rangle_E| $, where $x,y \in \mathbb{R}^d$.
This section is dedicated to verifying these hypotheses for the sequence (\ref{SN}).

Let $T_\mu$ be the smallest closed semigroup in $\mathrm{Gl}(d, \mathbb{R})$, which contains the support of a probability measure $\mu$ on $\mathrm{Gl}(d, \mathbb{R})$. The next theorem is due to Le Page and was improved by Lacroix, see Corollary 2.3 in \cite[Ch.VI]{BoLa85}. 
\begin{theorem}\label{CLth}
Suppose that the semigroup $T_\mu$ is strongly irreducible and contracting and that, for some $\tau >0$,
$\int e^{\tau l(\widehat M)} d\mu(\widehat M) < \infty$, where $l(\widehat M)=\sup \{\ln^+ \Vert \widehat M \Vert, \ln^{-}\Vert \widehat M \Vert)\}$,
with $\widehat M \in M =\{B, R\}$ and
$\Vert \widehat{M} \Vert =\sup \{ \Vert \widehat{M}x \Vert : x \in \mathbb{R}^3, \Vert x \Vert=1 \}$.
Let us denote by $\alpha$ the upper Lyapunov exponent\footnote{here, the upper Lypunov exponent is defined as $\alpha=\lim_{n\rightarrow \infty}\mathbb{E}\left[\frac{1}{n}\ln \|S_n\|\right]$.}
associated with $(S_n)_{n\geq 1}$.
Then there is a constant $\beta >0$ such that
\begin{itemize}
\item[(i)] For all non-zero vectors $x,y\in \mathbb{R}^d$,
$$
\lim_{n\rightarrow \infty } \frac{1}{n} \ln |\langle S_n x,y\rangle_E | = \alpha,
$$
and $\frac{1}{\beta\sqrt{n}}\big(\ln |\langle S_n x,y\rangle_E |-n\alpha \big)$ converges in distribution to the Gaussian law $\mathcal{N}(0,1)$.
\item[(ii)] If $(S_n)_{ij},\; 1\leq i,j \leq d, $ are the coefficients of the matrix $S_n$, then the $\mathbb{R}^{d^2}$-valued random vector
$$
\frac{1}{\beta\sqrt{n}}\big(\ln |(S_n)_{ij} | -n\alpha \big),\quad 1\leq i,j \leq d,
$$
converges in distribution to a random vector $Y$ which satisfies $Y_{ij}=Y_{kl}$ for all $1\leq i,j,k,l\leq d$ and such that $Y_{11}$ is distributed according to $\mathcal{N}(0,1).$
\end{itemize}
\end{theorem}
\begin{remark}
Combining the results from \cite[Theorem 2]{LePa82} and \cite[Part A,Ch.VI, Proposition 2.2]{BoLa85} it follows readily that
$$
\alpha = \lim_{n\rightarrow \infty}\mathbb{E} \left[\frac{1}{n}\ln |\langle S_n x,y\rangle_E| \right],
$$
and 
$$
\beta^2 = \lim_{n\rightarrow \infty}\mathbb{E}\left[\frac{1}{n}(\ln |\langle S_n x,y\rangle_E|-n\alpha)^2\right],
$$
both limits being independent of the non-zero vectors $x,y\in\mathbb{R}^d$. 
\end{remark}
Later on we shall apply the previous central limit theorem and for this we need to verify strong irreducibility and contractivity of the semigroup $T_\eta$ in $\mathrm{Gl}(3, \mathbb{R})$. Note that, since $\eta$ is finite, the third condition is automatically fulfilled.

\medskip

\noindent
{\it Definition of strong irreducibility}.
A subset $S$ of $\mathrm{Gl}(d,\mathbb{R})$ is called strongly irreducible if there does not exist a finite family of proper subspaces of $\mathbb{R}^d$, $V_1, V_2,\ldots, V_l$, so that for all $\widehat M \in S$
$$\widehat M(V_1 \cup V_2 \cup \ldots \cup V_l) = V_1 \cup V_2 \cup \ldots \cup V_l.$$
\medskip
\noindent
{\it Equivalent characterization of strong irreducibility.}
The semigroup $T_\mu$ is strongly \\[-1.2ex] irreducible if and only if $\text{supp}(\mu)$ is strongly irreducible, see \cite[p.48]{BoLa85}. \\[0.5ex]
In order to prove strong irreducibility of $T_\eta$, we shall make use of Lemma \ref{stirred}, taken from \cite[Part A, Ch.III]{BoLa85}:

\begin{lemma}\label{stirred}
If $T_\mu$ is not strongly irreducible one can find subspaces $V_1, V_2,\ldots, V_l$ with the same dimension and the properties
\begin{itemize}
  \item [a)]\text{If} $i \neq j$, $V_i \cap V_j = \{0\}$;
  \item [b)]$\forall i \in \{1,2,\ldots,l\}$ and $\widehat M \in M$, $\widehat M V_i = V_j$ for some $j \in \{1,2,\ldots,l\}$.
\end{itemize}
\end{lemma}
In the present setting, proving strong irreducibility amounts to showing the following lemma
\begin{lemma}\label{strongirred}
For every family of subspaces $V_1, V_2,\ldots, V_l$ with the same dimension $1$ or $2$ we have:
\begin{eqnarray}
  &\text{a)}&  \text{If}\; i \neq j,\, V_i \cap V_j = \{0\}; \nonumber \\
  &\text{b)}&
 \text{there is some}\; i \in \{1,2,\ldots,l\} \; \text{and} \; \widehat M \in M \text{\,so that}\; \widehat M V_i \neq V_j, \nonumber \\
&& \text{for all}\;  j \in \{1,2,\ldots,l\}. \label{strongirreda}
\end{eqnarray}
\end{lemma}

Before proving Lemma \ref{strongirred}, we verify in the next lemma, under specific hypotheses, property (\ref{strongirreda}) in the one-dimensional case:

\begin{lemma}\label{strongirred1}
Let $V_1= \{k_1 \underline{v}\}$ be the first proper subspace in a sequence of one-dimensional subspaces $V_1,\ldots,V_l$. If the following inequalities
\begin{align}
B^j(RB)^n \underline{v} \; &\neq \,  K B^s(RB)^m \underline{v}\,,\label{irreda} \\
R^j(BR)^n \underline{v} \; &\neq \,  K R^s(BR)^m \underline{v}
\end{align}
hold, for all $j,s \in \{0,1\}$, for all $n\geq m$, $n,m \in \mathbb{N}$, and every $K \in \mathbb{R}$, property
(\ref{strongirreda}) is satisfied in the one-dimensional case.
\end{lemma}

\medskip
\noindent
{\bf Proof of Lemma \ref{strongirred1}:} Consider the vector $B \underline{v}$. It does not belong to the subspace $V_1$ by hypothesis (\ref{irreda}), choosing $j=1, s=0, n=m=0$. Hence one has that $B\underline{v} \notin V_i$, $\forall i \in \{1,2,\ldots,l\}$ or $B\underline{v} \in V_i$, for some $2 \leq i \leq l$. In the first case the lemma is proved, otherwise there exists an $i$, for instance $i=2$, such that $B\underline{v} \in V_2 = \{k_2 B \underline{v}\}$. We repeat the above procedure, by applying alternatively the matrices $R$ and $B$.
After  $t$ steps we obtain the vector $B^j(RB)^{\frac{t-j}{2}}\underline{v}$,
where $j=0$ for $t$ even and $j=1$ for $t$ odd. Let $\overline{t}$ be the maximum index
such that for any $t \leq \overline{t}$ the vector $B^j(RB)^{\frac{t-j}{2}}\underline{v}$ did not exit from $V_1, V_2, \ldots, V_l$. It means that $V_1=\{k_1 \underline{v}\}$, $V_2=\{k_2 B \underline{v}\}$,
$V_3=\{k_3 RB \underline{v}\},\ldots,V_{\overline{t}+1}=\{k_{\overline{t}+1} B^j (RB)^{\frac{\overline{t}-j}{2}} \underline{v}\}$ and $V_{\overline{t}+2}$ is different from all of $V_1,\ldots,V_l$. If $\overline{t} < l-1$ the lemma is proved, otherwise for $\overline{t}=l-1$ we apply again $R$ or $B$ (depending on whether $t$ is even or odd), thereby obtaining the vector $(RB)^{\frac{l}{2}}\underline{v}$ for $l$ even and $B(RB)^{\frac{l-1}{2}}\underline{v}$ for $l$ odd. Such a vector does not belong to any $V_1, V_2,\ldots, V_l$, because of (3.2). The remaining case (3.3) is symmetric. The lemma is so proved.
\hfill{$\Box$}

\medskip

\noindent{\bf Proof of Lemma \ref{strongirred} in the one-dimensional case:}
As in the previous lemma let $V_1=\{k_1 \underline{v}\}$. In order to complete the prove in the one-dimensional case there remains to consider the following cases (\ref{nr32a}) or (\ref{nr32b}):
there exist $K \in \mathbb{R}$, $j,s \in \{0,1\}$, $n\geq m$, $n,m \in \mathbb{N}$, such that
 \bea
\label{nr32}
B^j(RB)^n \underline{v}\; =\, K B^s(RB)^m \underline{v}\, , \label{nr32a} \\
R^j(BR)^n \underline{v}\; =\, K R^s(BR)^m \underline{v}\,. \label{nr32b}
\eea
Consider first the case (\ref{nr32}) for $j=s \in \{0,1\}$. The second case (3.5) is symmetric.
For concreteness suppose that $j=s=1$. The other case can be treated analogously.
From (\ref{nr32}) we deduce that $(RB)^{n-m}\underline{v}=K \underline{v}$, because $B,R$ are invertible matrices. From Remark 2.1 and by Proposition 2.1 it follows that $K>0$ (eigenvalue of $(RB)^{n-m}$) and that $RB\underline{v}=\sqrt[n-m]{K} \underline{v}$, that is, $\underline{v}$ is an eigenvector of $RB$ and $\sqrt[n-m]{K} $ the corresponding eigenvalue. By applying the matrix $B$ to the vector $\underline{v}$ we get the vector $B\underline{v}$, which cannot belong to the subspace $V_1$, because otherwise $\underline{v}$ would be also an eigenvector of $B$, and hence, an eigenvector of $R$. This is not possible, because the eigenvectors of $B$ are distinct from those of $R$, for every $(u,v,w)$.
Therefore, there remain the following possibilities:
\begin{itemize}
  \item [(i)] $B\underline{v} \notin V_i$, for any $i \in \{1,2,\ldots,l\}$\,,
  \item [(ii)] $\exists i \in \{2,\ldots,l\}$ such that $ B\underline{v} \in V_i$\,.
\end{itemize}
In case (i) we are done, otherwise, there exists an $i$, for instance $i=2$, such that $V_2=\{k_2 B\underline{v}\}$.
Let us iterate this procedure as in the previous lemma, by considering the vector $B^2\underline{v}$, and so on.
After  $t$ steps we obtain the vector $B^t\underline{v}$. Let $\overline{t}$ be the maximum index such that for any $t \leq \overline{t}$ the vector $B^t \underline{v}$ does not exit from $V_1, V_2, \ldots, V_l$. It means that $V_1=\{k_1 \underline{v}\}$, $V_2=\{k_2 B \underline{v}\}$,
$V_3=\{k_3 B^2 \underline{v}\}$,\ldots, $V_{\overline{t}+1}=\{k_{\overline{t}+1} B^{\overline{t}} \underline{v}\}$. If $\overline{t} < l-1$, the lemma is proved, otherwise for $\overline{t}=l-1$ we apply again the matrix $B$, thereby obtaining the vector $B^l\underline{v}$.
The vector $B^{l} \underline{v}$ does not belong to any $V_i$, for any $1 \leq i \leq l$, because if $B^{l} \underline{v} \in V_i$, for some $i$, then $\underline{v}$ is an eigenvector of $B$ and $R$ at the same time, which is not possible.
The Lemma is so proved under the assumption (3.4) for $j=s$.

Let us assume that $j=1$ and $s=0$ in formula (3.4). The other case, $j=0$ and $s=1$, is symmetric.
Suppose there exist $K \in \mathbb{R}$ and $n \geq m$ such that:
\be
\label{nr33}
B(RB)^{n-m}\underline{v}= K \underline{v}\, .
\ee
The vector $RB\underline{v}$ does not belong to the subspace $V_1$, otherwise $\underline{v}$ would be an eigenvector of $RB$. By (\ref{nr33}), $\underline{v}$ would also be an eigenvector of $B$ and, hence of $R$, but this is not possible.
Therefore, we are left with the following alternatives:
\begin{itemize}
  \item [(i)] $RB\underline{v} \notin V_i$, for any $i \in \{1,2,\ldots,l\}$\,,
  \item [(ii)] $\exists i \in \{2,\ldots,l\}$ such that $ RB\underline{v} \in V_i$\,.
\end{itemize}
If case (i) occurs, the Lemma is proved, otherwise, there exists an $i$, for instance $i=2$, such that $V_2=\{k_2 R B\underline{v}\}$.
As above, after  $t$ steps we obtain the vector $(RB)^t \underline{v}$. Let $\overline{t}$ be the maximum index such that for any $t \leq \overline{t}$ the vector $(RB)^{t}\underline{v}$ does not exit from $V_1, V_2,\ldots, V_l$. It means that $V_1=\{k_1 \underline{v}\}$, $V_2=\{k_2 RB \underline{v}\}$,
$V_3=\{k_3 (RB)^2 \underline{v}\}$,\ldots, $V_{\overline{t}+1}=\{k_{\overline{t}+1}  (RB)^{\overline{t}} \underline{v}\}$. If $\overline{t} < l-1$, the lemma is proved, otherwise for $\overline{t}=l-1$ we apply once more $RB$, thereby obtaining the vector $(RB)^{l}\underline{v}$. Such
a vector does not belong to any $V_i$, for no $1 \leq i \leq l$, because if $(RB)^{l} \underline{v} \in V_i$, for some $i$, then $\underline{v}$ would be an eigenvector of $RB$. Moreover, by (\ref{nr33}), $\underline{v}$ would also be an eigenvector of $B$, and hence of $R$, which is not possible.
The lemma is thus proved in the one-dimensional case.
\hfill{$\Box$}
\medskip

\noindent{\bf Proof of Lemma \ref{strongirred} in the two-dimensional case:} In order to prove strong irreducibility in the two-dimensional case
we have only to show that for any proper subspace $V$ of dimension $2$ there exists a vector $\underline{v}\in V$ such that $B\underline{v}\notin V$ or $R\underline{v}\notin V$.
This is because now we have only one proper subspace, $l=1$ within property (\ref{strongirreda}), for otherwise the condition $V_i \cap V_j = \{0\}$ is not satisfied.

Let $\underline{n}$ be the vector normal to the plane $V$, expressed in polar coordinates as:
$\underline{n}=(\sin{\theta} \cos{\varphi}, \sin{\theta} \sin{\varphi}, \cos{\theta})$, with $\theta \in [0, \pi]$ and $\varphi \in [0, 2\pi)$.
Let $\{ \underline{v_1}, \underline{v_2}\}$ be the following orthonormal basis on the plane $V$:
$\underline{v_1}=(\cos{\theta} \cos{\varphi}, \cos{\theta} \sin{\varphi}, - \sin{\theta})$,
$\underline{v_2}=(- \sin{\varphi},  \cos{\varphi}, 0)$. Then the triple $B_1=(\underline{v_1}, \underline{v_2}, \underline{n})$ forms an orthonormal basis of $\mathbb{R}^3$.
Let $B_{B_1}=\{\tilde{b}_{ij} \}_{1 \leq i,j \leq 3}$ be the matrix $B$ with respect to the basis $B_1$.
One has the following identity: $B_{B_1}=M B M^{-1}$, where $M$ is the rotation matrix
$$M=\left(
\begin{array}{ccc}
\cos{\theta} \cos{\varphi} & \cos{\theta} \sin{\varphi} & - \sin{\theta} \\
-  \sin{\varphi} &  \cos{\varphi} & 0 \\
\sin{\theta} \cos{\varphi} & \sin{\theta} \sin{\varphi} & \cos{\theta}  \\
\end{array}
\right).$$
Let $R_{B_1}=\{\tilde{r}_{ij} \}_{1 \leq i,j \leq 3}$ be the matrix $R$ with respect to the basis $B_1$.
One has the following identity: $R_{B_1}=M R M^{-1}$.
Suppose we had $B \underline{v} \in V$ and $R\underline{v} \in V$ for every $\underline{v} \in V$.
This would entail the following equalities
$$\left\{
  \begin{array}{ll}
    \tilde{b}_{31}=0  \\
     \tilde{b}_{32}=0  \\
     \tilde{r}_{31}=0  \\
     \tilde{r}_{32}=0 \, .
  \end{array}
\right.
$$

Taking into account the definition of the matrices $B_{B_1}, R_{B_1}$, we get the following system
of equations
{\setstretch{1.3}
$$\left\{
  \begin{array}{ll}
    \tan^2{\varphi} - \frac{1}{u}\tan{\varphi}+\frac{1}{u} =0  \\
     \tan^2{\varphi} - \frac{1-u}{w}\tan{\varphi}+\frac{1-w}{w} =0   \\
     \tan^2{\varphi} - \frac{w(1-w)}{vw} =0   \\
     \cot{\theta}=\frac{1-v}{w} \cos{\varphi}\,.
  \end{array}
\right. $$}
The above system in turn implies a set of constraints as follows
{\setstretch{1.3}
$$\left\{
  \begin{array}{ll}
     \frac{1}{u}=\frac{1-u}{w}   \\
     \frac{1-w}{w}= \frac{w(1-w)-v}{vw} =0 \\
     \frac{1}{u} =-\frac{1-w}{w}\,.
  \end{array}
\right. $$}
It is easy to see that the above system does not admit any real solution $(\theta, \varphi)$ for any $(u,v,w)$. Hence
$R V \neq V$ or $B V \neq V$. The proof of strong irreducibility in the two-dimensional case is thus complete.
\hfill{$\Box$}

\medskip
For the central limit theorem being applicable, we still need to verify contractivity.\\[1ex]
{\it Definition:}
Given a set $T$ of $\mathrm{Gl}(d,\mathbb{R})$, one defines the index of $T$ as the least integer $p$ such that there exists a sequence $(M_n)_{n\geq 1}$
in $T$ for which  $\Vert M_n \Vert^{-1} M_n$ converges to a rank $p$ matrix. One says that $T$ is contracting when the index is $1$.
\medskip

In order to prove contractivity of the semigroup $T_\eta$ we apply the following property, see exercise $1.9$ in \cite[Part A, Ch.III]{BoLa85}:
\medskip

\noindent
{\it Sufficient condition for contractivity:} Suppose that a semigroup $T$ in $\mathrm{Gl}(3, \mathbb{R})$ contains a matrix $M$ with a unique eigenvalue of maximum modulus, this eigenvalue being simple. Then $T$ is contracting. \\[1ex]
{\it Region of contractivity:} In the present case, from Proposition $2.1$ we know that for every $u> \frac{1}{4}$ the above property is satisfied by the matrix $B$  for any $w^2 >u$ and any $v$. In the case where  $u< \frac{1}{4}$, contractivity holds for any $u \neq w(1-w)$ and any $v$. Finally, for $
u= \frac{1}{4}$ the above property holds for any $w>  \frac{1}{2}$ and any $v$.
\noindent
Symmetrically, for every $v> \frac{1}{4}$, the above property is satisfied by the matrix $B$  for any $w^2 >v$ and any $u$.  In the case where  $v< \frac{1}{4}$, contractivity holds for any $v \neq w(1-w)$ and any $u$. Finally, for $
v= \frac{1}{4}$ the latter property holds for any $w>  \frac{1}{2}$ and any $u$.
Note that at the critical point contractivity is not satisfied.

\begin{remark}
Let $C$ be the set where contractivity holds, more precisely $C=\{(u,v,w) \in (0,1)^3 :\text{contractivity holds for }T_{\eta}\}$.
\end{remark}
\section{Mean of the hard-dimer generating function}
 \setcounter{secnum}{\value{section}
 \setcounter{equation}{0}
 \renewcommand{\theequation}{\mbox{\arabic{secnum}.\arabic{equation}}}}
Let us recall formula (\ref{nr05}) from above
\be
\label{nr5} \langle Z_{\textnormal{hdc's on}\: \xi_N}(-u,-v,w)\rangle= 1+
\sum\limits_{k=1}^{{N-1}} \; \sum\limits_{s=1}^{k\wedge(N-k)}    \binom{k}{s} \binom{N-k-1}{s-1} \left (-\frac{u+v}{2w} \right )^s \left (\frac{w}{2}\right)^{N-k}.
\ee
The set $A$ of parameters $(u,v,w)$ where the series (\ref{nr5}) converges absolutely is given by
  \be
  \label{nr6}
  A= \left \{(u,v,w): \frac{u+v}{2} \leq w \leq 2 \right \}.
  \ee
Moreover, let $B$ be the set defined as
\be
\label{nr7}
B= \left \{(u,v,w)\in A: {u+v} <\left ( 1- \frac{w}{2}\right )^2  \right \}.
\ee

\begin{figure}[h]
\captionsetup{width=0.8\textwidth}
\begin{center}
\includegraphics[scale=0.63]{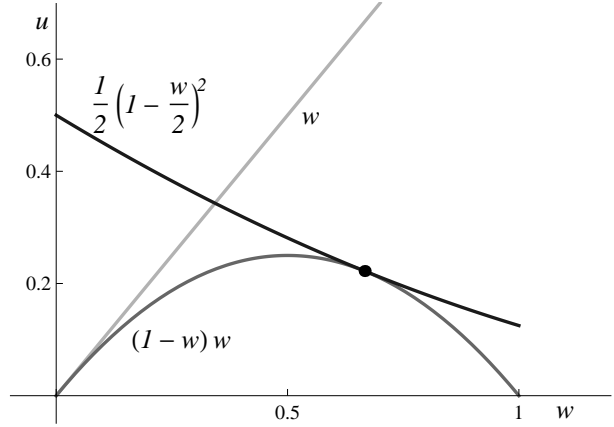}  
\end{center}
\caption{\small
The domain $A \cap B \cap C$ is shown for $u=v$. It is the region lying below the curves $u=w$ and $u=\frac{1}{2}\left(1-\frac{w}{2}\right)^2$. The dot marks the critical point. The curve $u=w(1-w)$ indicates the points where contractivity of $T_\eta$ is not known to hold.}
\label{ABC}
\end{figure}

The interesting feature of the next result is that we find for each $(u,v,w) \in A \cap B \cap C$ an exact formula of the averaged generating function.
Note that $A \cap B \cap C \neq \emptyset$, see also figure \ref{ABC}.
\begin{theorem}\label{thZN}
For each $(u,v,w) \in A \cap B \cap C$ the following formula
\be
\label{nr8}
\langle Z_{\textnormal{hdc's on}\: \xi_N}(-u,-v,w)\rangle =\frac{\tilde{C}}{S} \left (\frac{1+S+\frac{w}{2}}{2} \right )^{N-1}
\ee
holds, where the constants $\tilde{C}\equiv \tilde{C}(u,v,w)$ and $S\equiv S(u,v,w)$, given as
\begin{align}
\tilde{C} &= \frac{1}{2}\left( S +1 -\frac{u+v+w}{2} \right), \\[1ex]
 S &=\sqrt{1-(1-2z)w+\frac{w^2}{4}}\,,
\end{align}
with $z=-\frac{u+v}{2w}$, do not depend on $N$.
\end{theorem}
\begin{remark}
Note that for $(u,v,w) \notin A \cap B \cap C$, formula (\ref{nr8}) does not make sense, because $1-(1-2z)w+\frac{w^2}{4} \leq 0$. Moreover, the constant $\tilde{C}$ is positive in $A \cap B \cap C$ and the exponential function decays with respect to  $N$.
\end{remark}
{\bf Proof:}
The combinatorial factor $ \binom{k}{s} \binom{N-k-1}{s-1} $ in (\ref{nr5}) defines a non standardized hypergeometric distribution with respect to the index $s \in \{1,2,...,k\wedge(N-k)\}$. It is well known that its generating function can be expressed in terms of the hypergeometric function $_2F_1$, see \cite{BaI85}, so that
\be
\label{nr9}
\sum\limits_{s=1}^{k\wedge(N-k)} \binom{k}{s} \binom{N-k-1}{s-1} z^s =z \, k
\, _2F_1(-k+1,-N+k+1;2,z)\,.
\ee
Therefore, the averaged generating function defined in (\ref{nr5}) becomes
\be
\label{nr10}
\langle Z_{\textnormal{hdc's on}\: \xi_N}(-u,-v,w)\rangle = 1+ z \sum\limits_{k=1}^{N-1} (N-k)
\; _2F_1(-k+1,-N+k+1;2,z) \left( \frac{w}{2}\right )^k .
\ee
In (\ref{nr10}) we applied the property $ _2F_1(a,b;c,z)={}_2F_1(b,a;c,z)$. Note that for the present model the hypergeometric function is a Jacobi polynomial.
Therefore, we may use an exact formula for the generating function associated with Jacobi polynomials, see formula (1) in section 2.5.1 of \cite{BaII85}. More generally,
\be
\label{nr11}
\sum\limits_{r=0}^\infty \frac{(c)_r}{r!} \; _2F_1(-r,r+a;c,z)
\left( \frac{w}{2}\right )^k =
\frac{1}{S} \left ( \frac{S+\frac{w}{2}-1}{wz}\right )^{c-1}
\left ( \frac{S+\frac{w}{2}+1}{2}\right )^{c-a},
\ee
where $S$ is defined in (4.6) and $(c)_r= c(c+1)\cdots(c+r-1)$ for $r=1,2,3,\ldots$  and $(c)_0=1$. In order to apply formula (\ref{nr11}) to verify expression (\ref{nr8}) we use a property of contiguous functions, see \cite[p.103]{BaI85},
\begin{align}
& \hspace{1cm} z \, (N-k)
\, _2F_1(-k+1,-N+k+1;2,z)= \nonumber \\[1ex]
& -(1-z) \, _2F_1(-k+1,-N+k+1;1,z) + \,_2F_1(-k,-N+k+1;1,z)\,.
\end{align}
Substituting expression (4.10) into (4.8), we get
\bea
\langle Z_{\textnormal{hdc's on}\: \xi_N}(-u,-v,w)\rangle =
\sum\limits_{k=0}^{N-1}
\; _2F_1(-k,-N+k+1;1,z) \left( \frac{w}{2}\right )^k  \nonumber \\ -\, (1-z) \; \frac{w}{2} \;
\sum\limits_{k=0}^{N-2}
\; _2F_1(-k,-N+k+2;1,z) \left( \frac{w}{2}\right )^k\,.
\eea
Applying formula (\ref{nr11}) to the first sum in (4.11), with $a=-N+1$, and to the second sum in (4.11), with $a=-N+2$, we obtain
\bea \label{ZNsplit}
\langle Z_{\textnormal{hdc's on}\: \xi_N}(-u,-v,w)\rangle) =
\frac{1}{S}
\left ( \frac{S+\frac{w}{2}+1}{2}\right )^{N} -
\frac{1}{S}\; (1-z) \; \frac{w}{2}
\left ( \frac{S+\frac{w}{2}+1}{2}\right )^{N-1}
\eea
and hence the thesys.
\hfill{$\Box$}
\section{Discussion and outlook}
\setcounter{secnum}{\value{section}
 \setcounter{equation}{0}
 \renewcommand{\theequation}{\mbox{\arabic{secnum}.\arabic{equation}}}}
In this section we discuss how the findings above can be exploited to better understand the gravity model of BeLoZa.

Note that
$$
(\mathcal{S}_N)_{11} = \langle \lambda,\mathcal{S}_N\gamma \rangle_E 
= \langle \mathcal{S}_N^t \lambda,\gamma \rangle_E = \langle M_N^t M_{N-1}^t\cdots M_1^t \lambda,\gamma \rangle_E \,.
$$
Since the matrices $B^t,R^t$ satisfy the assumptions of Theorem \ref{CLth}, the latter gives

$$
\lim_{N\rightarrow \infty} \frac{1}{\beta\sqrt{N}}(\ln (\mathcal{S}_N)_{11}-N\alpha)=\mathcal{N}(0,1),\quad \text{(in distribution)}
$$
or 
\be\label{SN11}
(\mathcal{S}_N)_{11}\asymp e^{\beta\sqrt{N}\,\mathcal{N}(0,1)\, +\,N\alpha}=
\ln\mathcal{N}(N\alpha,N \beta^2),\quad (\text{in distribution}).
\ee
Here $\ln\mathcal{N}(\mu,\sigma^2)$ denotes a log-normal random variable $Y$, i.e. $Y=e^X$, where $X$ is a Gaussian random variable with mean $\mu$ and variance $\sigma^2$.
The probability density function of $Y$ is given by 
\be
f(y)=
\begin{cases}
\frac{1}{\sqrt{2\pi}\sigma y}e^{-\frac{(\ln y-\mu)^2}{2\sigma^2}}, \quad y>0 \\[0.5ex]
0, \quad y\leq 0 
\end{cases}
\ee
and $\langle Y \rangle =e^{\mu+\frac{1}{2}\sigma^2}$, whereas $\Big< \displaystyle{\frac{1}{Y}} \Big> =e^{-\mu+\frac{1}{2}\sigma^2}$. 
Splitting $(\mathcal{S}_N)_{11}$ into a Gaussian part $G_N=e^{\beta\sqrt{N}\,\mathcal{N}(0,1)\, +\,N\alpha}$ and a correction part $Q_N$, i.e. $(\mathcal{S}_N)_{11}=G_N+Q_N$, we may write, neglecting the perturbation $Q_N$,
\be\label{mean}
\langle(\mathcal{S}_N)_{11}\rangle=\left<Z_{\textnormal{hdc's on}\: \xi_N}(-u,-v,w)\right>\asymp e^{N\alpha +\frac{1}{2}N\beta^2},
\ee
and for the inverse
\be\label{meaninverse}
\left< \frac{1}{Z_{\textnormal{hdc's on}\: \xi_N}(-u,-v,w)}\right>\asymp e^{N\beta^2}\frac{1}{\left<Z_{\textnormal{hdc's on}\: \xi_N}(-u,-v,w)\right>}\,.
\ee
Therefore, expression (\ref{meaninverse}) gives the asymptotic behaviour when fluctuations of order higher than two are neglected.
The reason for the splitting above is motivated by relation (\ref{SN11}). Note that it is not possible to deduce directly from (\ref{SN11}) any asymptotic characteristic for the means $\langle(\mathcal{S}_N)_{11}\rangle$ and $\langle(1/\mathcal{S}_N)_{11}\rangle$. This is because (\ref{SN11}) involves an $N$-dependent exponential increase or decay, depending on whether $\alpha$ and $\beta^2$ together give a positive or negative number. And even though the convergence in distribution is uniform, this does not suffice to control 
the integrals involved.

To see what kind of conclusions can be drawn from the central limit theorem, we shall, for the rest of this section assume that only the first two moments $\alpha$ and $\beta^2$ are relevant.
Then, owing to (\ref{mean}) and (\ref{meaninverse}), $Z(x,y,\Delta t=1)$ depends on $(u,v,w)$ through $\alpha,\beta^2$ only. On the other hand, $\alpha,\beta^2$ can be expressed equally well through the Lyapunov exponents $L_1,L_2$, as defined in \cite{BeLoZa07}, which are more accessible in terms of concrete values
\be\label{char1}
L_1=\lim_{N\rightarrow \infty} \frac{1}{N}\ln \langle (\mathcal{S}_N)_{11}\rangle = \ln \nu_1,
\ee
and
\be\label{char2}
L_2=\lim_{N\rightarrow \infty} \frac{1}{N}\ln \langle (\mathcal{S}_N)_{11}^2\rangle = \ln \nu_2,
\ee
with $\nu_1,\nu_2$ being the largest eigenvalues of $1/2(B+R)$ and $1/2(B\otimes B + R\otimes R)$, respectively. In fact, it is not difficult to see that 
$$
\langle (\mathcal{S}_N)_{11}^2\rangle = \langle \lambda\otimes\lambda, \frac{1}{2}(B\otimes B + R\otimes R)^N \gamma\otimes\gamma\rangle_E.
$$
The relation to the parameters $\alpha,\beta^2$ appearing in the central limit theorem above can be seen from
$$
\langle (\mathcal{S}_N)_{11}\rangle\asymp e^{N\alpha+\frac{1}{2}N \beta^2}, \quad
\langle (\mathcal{S}_N)_{11}^2\rangle\asymp e^{2 N\alpha + 2 N \beta^2},
$$
giving 
\be\label{char3}
\alpha= 2L_1-\frac{L_2}{2},\quad \beta^2=-2L_1 + L_2.
\ee
Let us try to sum up the series (\ref{nr21}). To see when the latter converges we insert the asymptotic relation (\ref{meaninverse}), giving
\bea\label{sum}
\overline{Z}(x,y,\Delta t=1)&=& \sum_N e^{N(\ln 2-\gamma)} e^{N \beta^2}\frac{1}{\left<Z_{\textnormal{hdc's on}\: \xi_N}(-u,-v,w)\right>}\,. 
\eea 
If we consider (\ref{sum}) merely as a mathematical object, then convergence of (\ref{sum}) will depend on whether $\gamma$ is large enough to compensate the other two divergent factors. In fact, within the domain $A\cap B\cap C$ the variance $\beta^2$ is always a strictly positive number. In Appendix \ref{appendix} it is shown that the maxima of the factor $1/\left<Z_{\textnormal{hdc's on}\: \xi_N}(-u,-v,w)\right>$ grow exponentially fast, too. A closer look at the behaviour of the latter expression reveals that the arguments of its maxima lie within the domain $A\cap B\cap C$ and converge to the critical point $(u_c,v_c,w_c)$ as $N\rightarrow \infty$, cf. Appendix \ref{appendix}. Moreover, numerical computations show that on the same domain $\beta^2(u,v,w)$ satisfies $\beta^2(u',v',w)<\beta^2(u'',v'',w)$ for $u'<u''$ and $v'<v''$. Renormalizing the terms in (\ref{sum}) so that all their maxima are equal to 1, the series (\ref{sum}) can be turned into one which converges everywhere but at the critical point. This is because the terms then decay exponentially. The existence of at least one singular point is related to the feasibility of the continuum limit. In this limit one lets the length $a$ of edges in the triangulation tend to zero in such a manner that the physical quantities remain finite. For example, the physical space-time volume $\mathcal{V}$ should be finite and scale like $\langle \mathcal{V} \rangle=\lim_{a\rightarrow 0} a^3 \langle V \rangle,$ where the discrete space-time volume is given by (see \cite[eq.(61)]{BeLoZa07}
\be\label{discreteV}
\langle V \rangle = \left(-\frac{\partial}{\partial \lambda}\ln Z_N \right)=\left(2b_1 u \frac{\partial}{\partial u}+2 b_1 v \frac{\partial}{\partial v}+b_2 w\frac{\partial}{\partial w}\right)\ln Z_N,
\ee
where $\lambda$ is the bare cosmological constant and $b_1,b_2$ are constants related to the discrete geometry.
Therefore, $\langle V \rangle$ can become infinite only if a singularity is available.

A particularly neat expression is obtained if we use the asymptotic form
\be
\frac{1}{\left< Z_{\textnormal{hdc's on}\: \xi_N}(-u,-v,w)\right>} \sim \frac{1}{\nu_1^{N-1}},
\ee
where in addition the term $S$ was omitted. The sum (\ref{sum}) then simplifies to a geometric series
\bea\label{sum1}
\overline{\overline{Z}}(x,y,\Delta t=1)&=& \sum_N e^{N(\ln 2-\gamma)} e^{N \beta^2}\frac{1}{\nu_1^{N-1}} \nonumber \\
&=& \frac{\nu_1}{\tilde{C}}\sum_N e^{N(\ln 2-\gamma)}e^{N(-2L_1+L_2-\ln \nu_1)} \nonumber \\
&=& \frac{\nu_1}{\tilde{C}} \frac{1}{(1-e^{(\ln 2-\gamma) + \ln \nu_2-3\ln \nu_1})}.
\eea

The function $Z(x,y,\Delta t=1)$ not only encodes the random nature of the discrete geometry but it can also be used to perform the continuum limit. In \cite{BeLoZa07} it was shown that a canonical scaling is given by
\be\label{scaling}
u=\frac{2}{9}e^{-2a^2 X-2a^3b_1\Lambda},\quad v=\frac{2}{9}e^{-2a^2 Y-2a^3b_1\Lambda},\quad w=\frac{2}{3}e^{-a^3 b_2\Lambda}.
\ee
Furthermore, to make sure that (\ref{sum1}) diverges only at the critical point $(u_c,v_c,w_c)$ and converges otherwise, we have to replace $\gamma$ by $\gamma'=\ln 2+(-\ln \nu_2+3\ln \nu_1)(u_c,v_c,w_c)$. Finally, inserting the explicit expression of $\nu_1$ in (\ref{sum1}) and the perturbative expression \cite[eq.(85)]{BeLoZa07} for $L_2$ (and therefore for $\nu_2$), we get
\begin{gather}
  a^2 \overline{\overline{Z}}(X,-Y;\Lambda)= \nonumber \\
a\frac{6}{\sqrt{X-Y}} + a^2\left(-2+3\frac{XY}{(X-Y)^2}-\frac{\Lambda}{(X-Y)^{3/2}}+\frac{1}{12(X-Y)^{7/2}}\right)+O(a^3).
\end{gather}
Following the same line of arguments as in BeLoZa, one finds the continuum Hamiltonian
\be\label{Hamiltonian}
\overline{\overline{H}}_{\mathcal{A}}=C(-\mathcal{A}^{3/2}\frac{\partial^2}{\partial \mathcal{A}^2}-\frac{3}{2}\mathcal{A}^{1/2}\frac{\partial}{\partial \mathcal{A}}-\frac{1}{16}\frac{1}{\mathcal{A}^{1/2}}+\Lambda \mathcal{A}+\mathrm{const}\mathcal{A}^3),\\[1ex]
\ee
which agrees with the Hamiltonian in BeLoZa apart from the constant $C=3$ and the last potential term.

Based on the assumption that the central limit theorem gives the core information of the model, we derived results which are physically reasonable. It would be desirable to know whether these results continue to hold when higher order fluctuations, or what amounts to the same thing, higher order Lyapunov exponents are taken into account. Therefore, it would be an interesting task for the future to find out analytically, how far it is legitimate to neglect the perturbation $Q_N$.
At the same time one should look out for other models in the realm of quantum gravity to see whether the central limit theorem provides the main clue for the understanding of asymptotics in a more general context.
\begin{appendix}
\section{Divergence behaviour of the inverse means}\label{appendix}

\begin{figure}[h]
\captionsetup{width=0.85\textwidth}
\begin{center}
\includegraphics[scale=0.65]{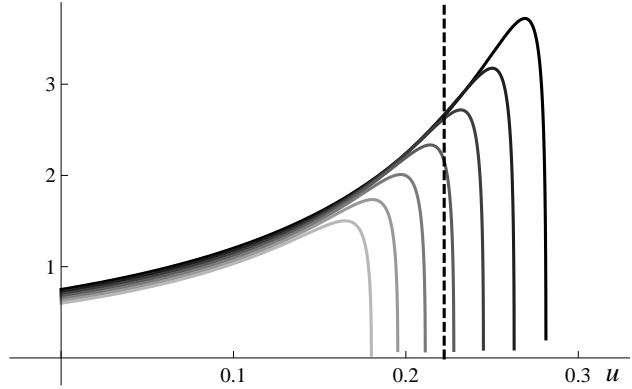}  
\end{center}
\caption{\small
This figure shows the family of curves for the function $2^{N-1}D_1(w)D_2(\,\cdot\,,w)$ for different $w$'s with $w\in\{0.5,0.55,0.6,0.65,2/3,0.7,0.75\}$ and $N=10$. The black curve corresponds to the smallest value $0.5$, and with increasing $w$'s the curves become brighter. The vertical line marks the value $u=2/9$.}
\label{fig3}
\end{figure}

Here we investigate the divergence of the inverse means in (\ref{sum}) exploiting equality (\ref{nr8}). Since $\tilde{C}$ just contributes a constant in the asymptotics we shall, to simplify the discussion, omit it. We therefore have to look at the behaviour of 
\be
\frac{1}{\langle Z_{\textnormal{hdc's on}\: \xi_N}(-u,-v,w)\rangle} \quad \text{as}\; 
N \rightarrow\infty. 
\ee
Using (\ref{nr8}), the latter can be cast in the following form
\be\label{newform}
\frac{S}{\nu_1^{N-1}}=
2^{N-1}
\underbrace{\left(1+\frac{w}{2}\right)^{-N+1}}_{D_1(w)}
\underbrace{S \left(1+\frac{S}{1+\frac{w}{2}} \right)^{-N+1}}_{D_2(u,w)}.
\ee
Elementary algebra shows that the factor $D_2$, when seen as a function of $u$, attains its maximum at
$$
u^\ast(N,w) = 
\frac{12-16N+4N^2-20w+16Nw-4N^2w+3w^2-4Nw^2+N^2w^2}{8(N-2)^2}.
$$
On the other hand, the function $u^\ast(N,w)$, seen as a function of $w$, is a convex polynomial for $N\geq 3$ attaining its minimum at $w=2(1+(2N+2)/(N^2-4N+3))$, which is always greater than the value $w=2/3$.
Let $w_{1,2}(N)$ be the two roots of the equation $u^\ast(N,w)= 2/9$, with $w_1(N)$ being the smaller one. More precisely,
$$
w_1(N)=\displaystyle{\frac{2(15-12N+3N^2-2\sqrt{48-64N+32N^2-8N^3+N^4})}{3(3-4N+N^2)}},
$$
which is well-defined for $N\geq 4$ and converges to $2/3$ from the left as $N\rightarrow \infty$.
For the rest of this paragraph we assume that $N\geq 4$ is fixed but arbitrary.
We observe that $u^\ast(N,\,\cdot\,)$ is decreasing on the interval $[0,2/3]$. This means that the value $u^\ast(N,w)$ is larger than $2/9$ for $w <w_1(N)$, and is smaller than $2/9$ for $w \geq w_1(N)$ and $w\in [0,2/3]$. Figure \ref{fig3} might help distinguish the different cases. Hence, the restriction of $D_2(\,\cdot\, ,w)$ to the domain $\mathcal{D}=\{u:0\leq u\leq 2/9\}$ assumes its maximum at $u^\triangle(N,w)=2/9$, for $w < w_1(N)$, and at
$u^\triangle(N,w)=u^\ast(N,w)<2/9$, for $w \geq w_1(N)$.
Therefore, on $\mathcal{D}$ the supremum of the function $D_1(w)D_2(\,\cdot\, ,w)$ equals
$D_1(w)D_2(2/9,w)$, when $w < w_1(N)$, and it equals $D_1(w)D_2(u^\ast(N,w),w)$, when $w \geq w_1(N)$.
In addition, the function $D_1(\,\cdot\,)D_2(u^\ast(N,\,\cdot\,),\,\cdot\,)$ is continuous and decreasing on $[0,1]$.
All these facts together imply that the function $D_1(w)D_2(u,w)|_{[0,\frac{2}{9}]\times[0,\frac{2}{3}]}$ assumes its maximum at $w=w_1(N),u=u^\ast(N,w_1(N))=2/9$.
Therefore, inserting the value of $w=w_1(N)$ into expression $D_1(w)D_2(2/9,w)$, one finds
\be\label{growth}
D_1(w_1(N))D_2(2/9,w_1(N))=\frac{1}{eN}\left(\frac{3}{4}\right)^{N-2}.
\ee
\end{appendix}

\end{document}